\documentclass[usenatbib]{HAWC2015}
\usepackage[numbers]{natbib}
\usepackage{graphicx}        
\newcommand{\prd}{Physical Review D}   

\begin{document}

\title*{IceCube at the Threshold}
\titlerunning{IceCube}
\author{T. Gaisser$^{1}$ for the IceCube Collaboration
}
\authorrunning{Gaisser} 

\institute{
$^{1}$Bartol Research Institute and Department of Physics and Astronomy,
University of Delaware, 
Newark, DE 19716\\
}
\maketitle
\vskip -3.8 cm  
\abstract{
IceCube has observed neutrinos above 100 TeV at a level significantly above the steeply 
falling background of atmospheric neutrinos.  The astrophysical signal is seen both in the 
high-energy starting event analysis from the whole sky and as a high-energy excess in the 
signal of neutrino-induced muons from below.  No individual neutrino source, either steady 
or transient, has yet been identified. Several follow-up efforts are currently in place in 
an effort to find coincidences with sources observed by optical, X-ray and gamma-ray detectors.  
This paper, presented at the inauguration of HAWC, reviews the main results of IceCube and describes 
the status of plans to move to near-real time publication of high-energy events by IceCube.
}
\vskip -1. cm
{\setlength{\unitlength}{1.cm}

\section{Introduction}

\vskip -0.2 cm

IceCube, the first kilometer-scale neutrino detector, 
was completed at the end of 2010 after seven Antarctic seasons.  The detector has
been in operation since May 2011 with its full complement of 86 strings of digital
optical modules (DOMs) in the deep ice
and 81 stations of the IceTop array on the surface (see Fig.~\ref{fig1:array}).  Somewhat like the relation
of HAWC to MILAGRO, the design of IceCube benefitted from its predecessor, AMANDA, as
described in the review by~\citet{AMANDA}.
In the case of IceCube, the main new feature of the design (apart from the greater size) 
is the data acquisition system (DAQ)
in which waveforms are digitized in each of the 60 DOMs on each string
and sent over copper cables to computers in the IceCube Lab for processing.  The ability to
drill with hot water, deploy optical modules in the deep ice 
and reconstruct neutrino-induced muons~\citep{2009PhRvD..79f2001A}
were all demonstrated in AMANDA.  The local digitization design of the DAQ was tested on String 18
of AMANDA~\citep{2006NIMPA.556..169A}.

Construction of
IceCube was supported by the U.S. National Science Foundation with funds from the Major
Research Equipment and Facilities Construction Account and by significant additional support 
from international partners.  It is interesting to note that construction of IceCube (which
by itself required transport of 4.7 million pounds of equipment to the South Pole)
proceeded in parallel with construction
of the new South Pole Station and of the South Pole Telescope.

\begin{figure}[ht]
\label{fig1:array}
  \centering
  \includegraphics[width=12 cm]{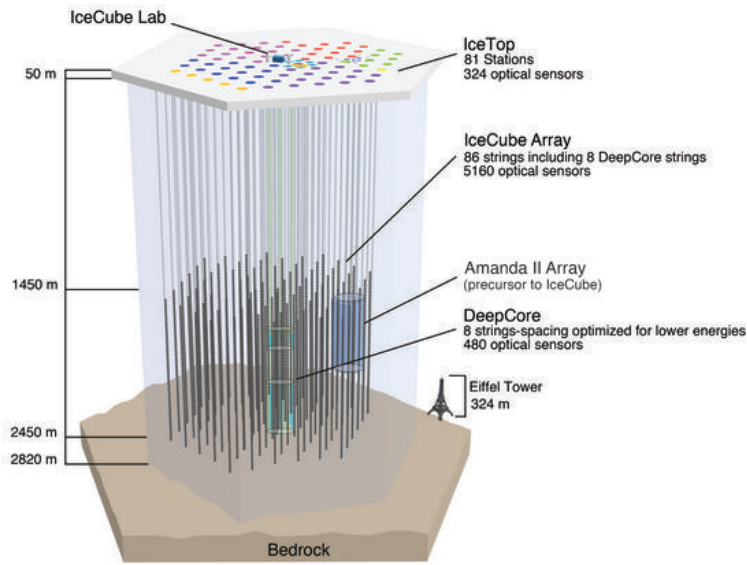}
  \caption{
Artist's view of the IceCube detector.
}
\end{figure}

Analysis of IceCube data
began soon after the first string and surface detectors were functioning in
2005-06~\citep{2006APh....26..155I} and continued with the first study
of neutrino-induced upward muons based on
data with nine strings in 2006-07~\citep{2007PhRvD..76b7101A}.  Studies
of neutrino-induced muons continued with increasing sensitivity.  They
provide the basis of searches for point sources of astrophysical 
neutrinos~\citep{2014arXiv1406.6757I} as well as a measurement of the spectrum
of atmospheric $\nu_\mu$~\citep{2011PhRvD..83a2001A}.  
The spectrum of atmospheric electron neutrinos in the TeV range was also
measured by making use of cascade-like events~\citep{2013PhRvL.110o1105A,2015arXiv150403753I}.
Closely related to the measurement of the flux of atmospheric $\nu_\mu$
is the search for astrophysical neutrinos, which would show up as a hardening
of the spectrum of muon neutrinos at high energy.  The measurement of upward $\nu_\mu$
with data from 2009-10~\citep{2013arXiv1311.7048T} included a few events with 
high energy, but not a statistically significant excess.

In 2012 two cascade-like
events with $\sim$PeV in deposited energy that started inside the detector were observed
in data taken in 2010-11 (IC 79) and 2011-2012 (IC 86).  The events were at
the lower boundary of the selection region of a search for much higher energy 
cosmogenic neutrinos
from photo-pion production on the cosmic background radiation~\citep{2013PhRvL.111b1103A}.
This discovery led to a dedicated search in the same two-year data sample for high energy
starting events (HESE).  Strong evidence for an excess of astrophysical neutrinos
at high energy above the steeply falling spectrum of atmospheric neutrinos
was found in a sample of 28 events that
included the two PeV events as well as a mixture of astrophysical
neutrinos and background in the 100 TeV range~\citep{2013Sci...342E...1I}.  
The discovery of astrophysical neutrinos 
was confirmed by a continuation of the HESE analysis to
a third year of data from 2012-2013 (IC 86-2), which included an event with
$\sim$2 PeV of deposited energy~\citep{2014PhRvL.113j1101A}.

The organization of this paper follows the order of the slides presented at Puebla, starting with
an overview of the broad scope of IceCube science.  Then some details of the
HESE analysis and related results are described.   This is followed by a section on
neutrino point source searches with IceCube and a discussion of implications
for what the sources of the astrophysical neutrinos observed by IceCube might be.
The paper concludes with a discussion of plans for the future, with emphasis
on the plan to make public in near real time information on the highest energy neutrinos.

\section{Overview of IceCube Science}

\noindent
{\bf Solar flares and supernova search:} The first 
extraterrestrial event seen by IceCube was the ground level cosmic-ray event
caused by the solar flare of December 13, 2006~\citep{2008ApJ...689L..65A}.
The event was seen as a sharp increase in the counting rate of the 32 tanks of IceTop
in operation at the time, as shown in Fig.~\ref{fig2:solar}(left).  

Scalar rates
of all DOMs in the deep detector are continuously monitored to look for
nearby supernovae~\citep{2011arXiv1108.0171I}.  The supernova signal would be 
caused by light generated by interactions of $\sim$10~MeV neutrinos within a few meters 
 of individual DOMs.

\vspace{0.2cm}
\noindent
{\bf Cosmic-ray physics with IceCube:} The IceTop air shower array consists of 81 stations,
each with two tanks separated by 10 m, with a spacing of approximately 125 m between
stations~\citep{2013NIMPA.700..188A}.  Each tank contains one DOM operating at high gain and one at low gain.  The
IceTop DOMs are fully integrated into the IceCube DAQ so that cosmic ray events seen
in coincidence by IceTop and the deep array of IceCube can be identified and reconstructed.
The all-particle spectrum measured with 73 stations of IceTop alone with data taken
in 2010-2011 is shown in the right panel of Fig.~\ref{fig2:solar}~\citep{2013arXiv1307.3795I}.
The conversion from the measured shower size parameter to primary energy depends
on primary composition.   The measurement
resolves significant structure in the spectrum, as indicated by the power-law fits
shown in limited energy ranges.

  \begin{figure}[t]
\label{fig2:solar}
  \centering
\includegraphics[width=5 cm]{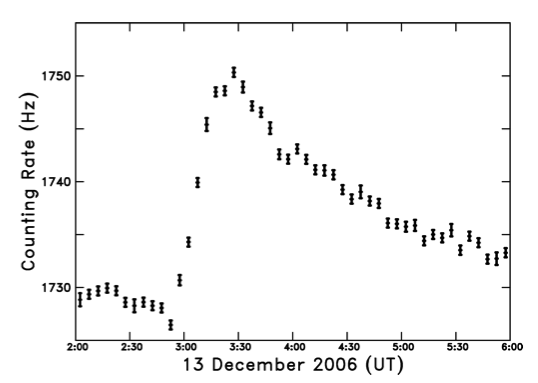}\,\,\includegraphics[width=6 cm]{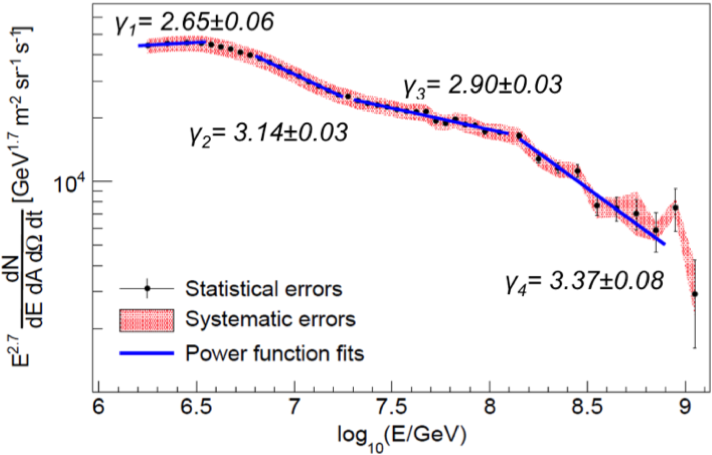}
  \caption{Left: Ground level solar flare event of December 13, 2006
  as seen in the 32 high-gain IceTop DOMs then in operation.  Right:
cosmic-ray spectrum from IceTop~\cite{2013arXiv1307.3795I} with power-law fits 
in four segments of the energy range.}
\end{figure}

  \begin{figure}[h]
  \centering
  \includegraphics[width=11cm]{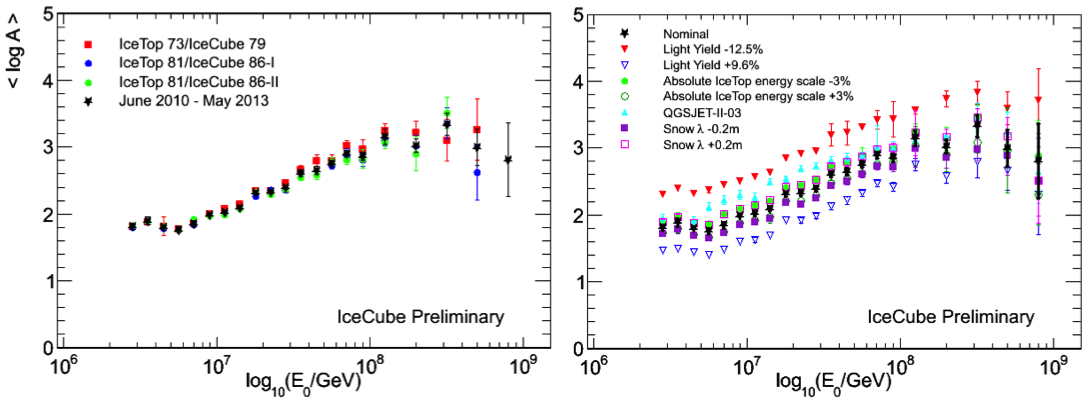}
  \caption{
Mean log of primary mass derived from three years of coincident events
(left, with statistical errors only; right, with systematic uncertainties).
}
\label{fig3:logA}
\end{figure}

It is also possible to select events seen in coincidence by IceTop and by the deep array
of IceCube as described in \citep{2013APh....42...15I,2013arXiv1309.7006I}.  
The ratio of signal from the muon bundle in the deep ice to the signal for
the shower at the surface is sensitive to the composition of the primary 
cosmic rays.  The analysis
uses a neural network with primary energy and primary mass as fitted variables.
An analysis is nearing completion on three years of coincident data consisting of
air showers well-reconstructed both in IceTop and in the deep array of IceCube.
  The preliminary result
is shown as a plot of the mean value of the natural logarithm of primary mass
in Fig.~\ref{fig3:logA}.  The energy spectrum obtained from this
independent analysis (which makes no a priori assumption about the primary
mass composition) confirms the IceTop only measurement (which does require an 
assumption about energy-dependence of the composition).  

A related analysis uses events reconstructed in IceTop with trajectories
pointing at the deep array of IceCube but with no muon hits in the deep detector.
Such muon poor events can be used to place
upper limits on PeV gamma rays from a region of the Galactic plane
at high negative declination.  An initial search based on data with half
the detector has been published~\citep{2013PhRvD..87f2002A}.

 \begin{figure}[th]
  \centering
  \includegraphics[width=10 cm]{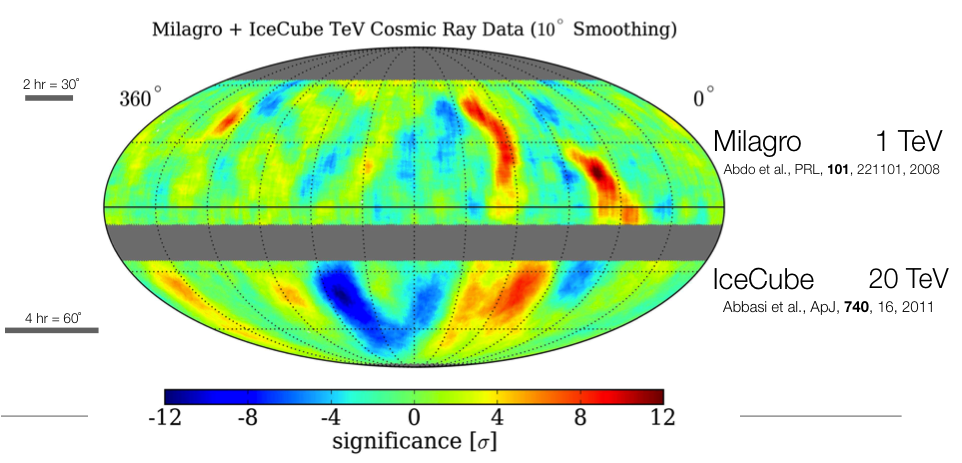}
  \caption{
Significance maps of cosmic-ray anisotropy measured by MILAGRO and IceCube.
}
\label{fig4:anisotropy}
\end{figure}

\vspace{0.2cm}
\noindent{\bf Cosmic-ray anisotropy with IceCube:} The event rate in IceCube is dominated by atmospheric muons from above with
 sufficient energy to penetrate to the detector ($>\approx 500$~GeV).  The rate
 in the full detector is about 3 kHz, with a $\pm$15\% seasonal variation.
 First pass directions and energies are assigned in the processing and filtering
 computers at the Pole.  
 These events have been used to measure the cosmic-ray anisotropy
 in the Southern sky for declinations $<-25^\circ$~\citep{2011ApJ...740...16A}.  A comparison with Northern
 hemisphere measurements by MILAGRO~\citep{2008PhRvL.101v1101A} is shown in Fig.~\ref{fig4:anisotropy}.
 A joint analysis between IceCube and HAWC will make a single map of the whole sky.
 
   \begin{figure}[hb]
  \centering
\includegraphics[width=5.5 cm]{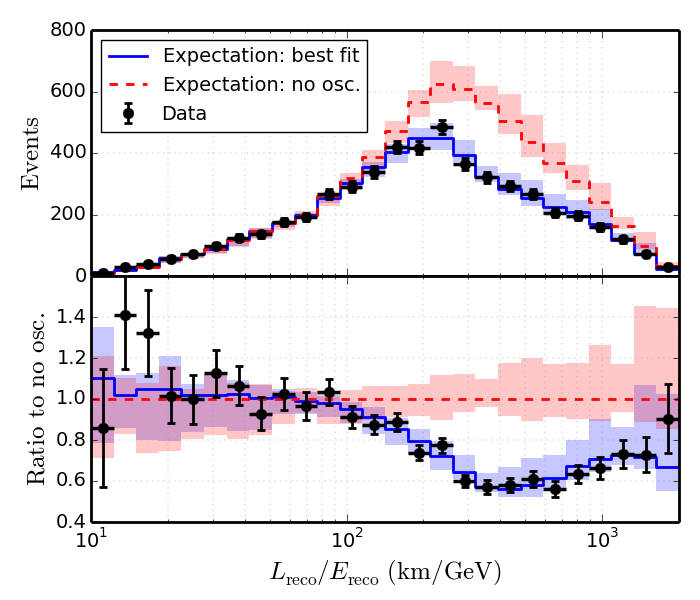}\,\,
\includegraphics[width=6 cm]{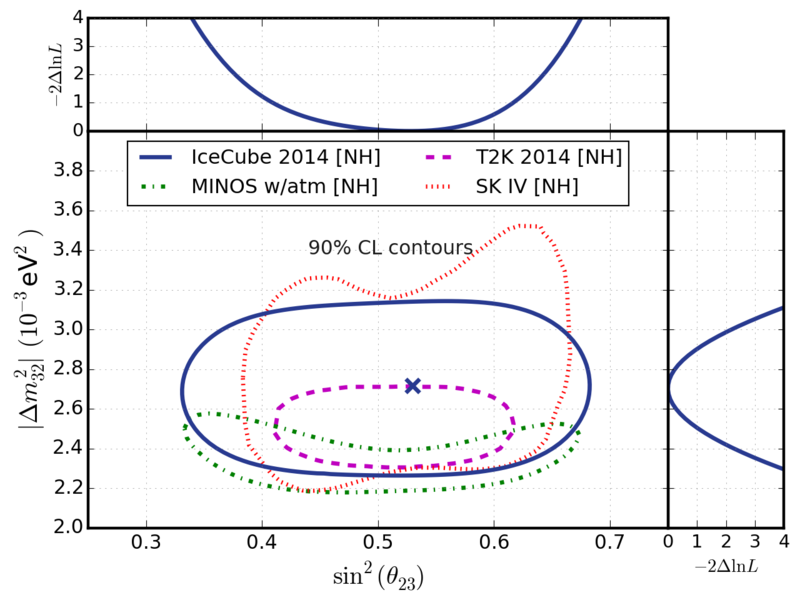}
  \caption{Left: Ratio ($L/E$) of path length through the Earth to neutrino energy 
  for atmospheric $\nu_\mu$ in IceCube.  Right:
Ninety \% confidence level contours for atmospheric neutrino oscillation parameters
from the IceCube analysis~\citet{2014arXiv1410.7227I}.}
\label{fig5:nuosc}
\end{figure}

\vspace{-0.2cm}
\noindent{\bf Neutrino oscillations:}
The closely spaced detectors of the DeepCore subarray of IceCube are used to select a
highly pure sample of low energy $\nu_\mu$ (6~-~56~GeV) 
from below that produce upward moving muons inside the detector.
Selection criteria are designed to ensure that the starting vertex and the decay point 
of each event are measured well.  The neutrino energy is then the sum of the
energy of the hadronic shower at the starting vertex and the muon range multiplied by
0.226~GeV/m.  Fits to the data are made with the 
physics quantities $\sin^2\theta_{23}$ and $\Delta m^2_{23}$ as free parameters.
The result is shown in Fig.~\ref{fig5:nuosc}~\citep{2014arXiv1410.7227I}.
A novel feature of this measurement is that the energy range
is dominated by the first oscillation minimum in the survival probability, 
$P_{\nu_\mu\nu_\mu}$ ($E_\nu\approx 25$~GeV).

  \begin{figure}[ht]
  \centering
\includegraphics[width=7.5 cm]{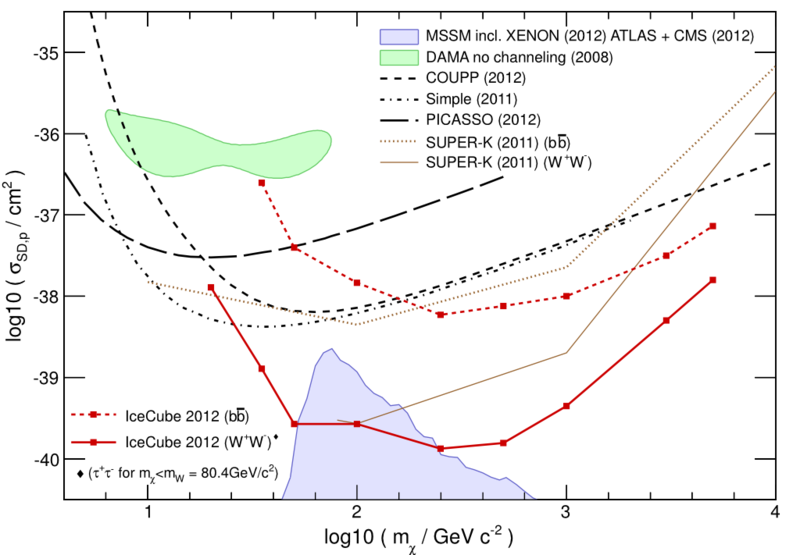}
  \caption{Upper limit on the WIMP-nucleon spin-dependent cross section based on
  a search for neutrinos from the Sun by~\citet{2012arXiv1212.4097I}.}
\label{fig6:solarWIMP}
\end{figure}
\noindent{\bf Search for dark matter:} There are several approaches to looking for neutrinos
from dark matter in IceCube.  The most sensitive is the upper limit from annihilation
of weakly interacting massive particles (WIMPs) after capture and concentration 
in the center of the Sun~\citep{2012arXiv1212.4097I}.
In equilibrium, the annihilation rate of WIMPs is equal to their capture rate.
Thus a limit on neutrinos from WIMP annihilation in the Sun is equivalent to
a limit on the capture cross section.
Because capture of WIMPs in the Sun is due largely to their interactions 
with hydrogen, the most significant limit is on the cross section for
spin-dependent interactions of WIMPs with nucleons, which is shown in 
Fig.~\ref{fig6:solarWIMP}.

  \begin{figure}[th]
  \centering
\includegraphics[width=5.0 cm]{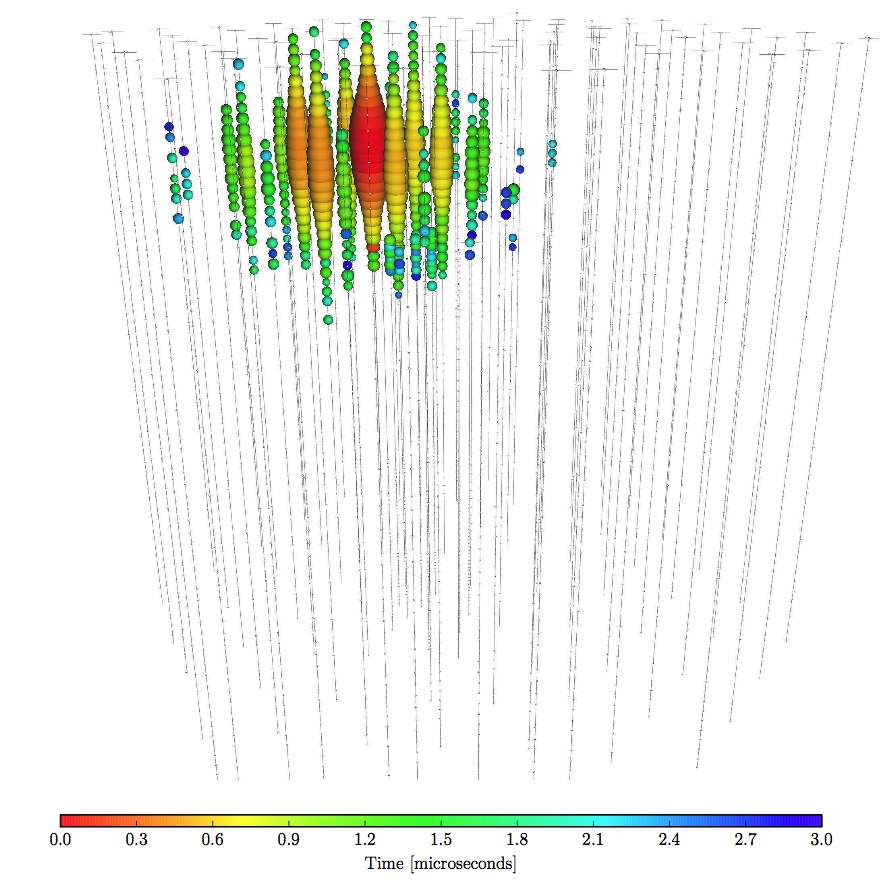}\,\,\,
\includegraphics[width=5.0 cm]{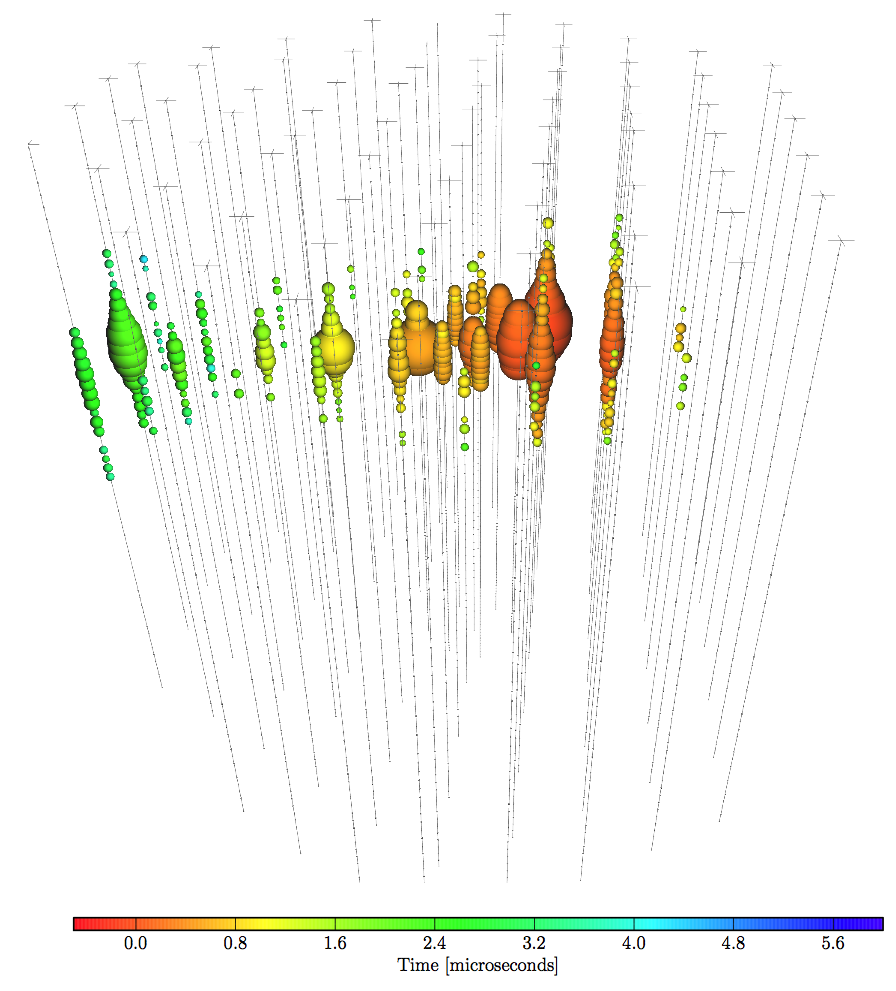}
  \caption{Left: Cascade event \#35 (2 PeV deposited energy);  Right:
Starting track event \#5 (70 TeV deposited energy).  Color code: red, early; green/blue late.}
\label{fig7:HESE-events}
\end{figure}

\vspace{-.6cm}
\section{Astrophysical neutrinos}
\vspace{-.2cm}

The key aspect of the HESE analysis is the criterion for selecting events
that start inside a fiducial volume surrounded by a veto region
and deposit a large amount of light in the detector. This strategy
has three key features. 
\begin{enumerate}
\item The analysis includes events from the whole sky. 
\item The background of atmospheric muons that pass the veto and
start inside the fiducial volume
can be determined from the data by studying muons tagged in the veto region
and observing how often they pass a suitably defined inner veto. 
\item Atmospheric neutrinos with sufficient energy so that a muon
produced in the same air shower would enter the detector are excluded
from the event sample.~\footnote{There are two cases of the atmospheric neutrino self-veto: 
when the vetoing muon is from the same decay in which the neutrino
was produced~\citep{2009PhRvD..79d3009S}; and a generalized veto that includes any
muon produced in the same air shower~\citep{2014PhRvD..90b3009G}.  
The latter is needed in particular
for electron neutrinos where the accompanying lepton at production is an electron
or positron.}

\end{enumerate}

Two events from the high energy starting event analysis are 
shown in Fig.~\ref{fig7:HESE-events}.
On the left is the highest energy neutrino observed, 
a cascade event with deposited energy of 2 PeV,
most likely from the charged current interaction of a $\nu_e$.  The track event on
the right starts inside the detector, deposits an estimated 70 TeV then leaves the
detector.

  \begin{figure}[th]
  \centering
\includegraphics[width=5 cm]{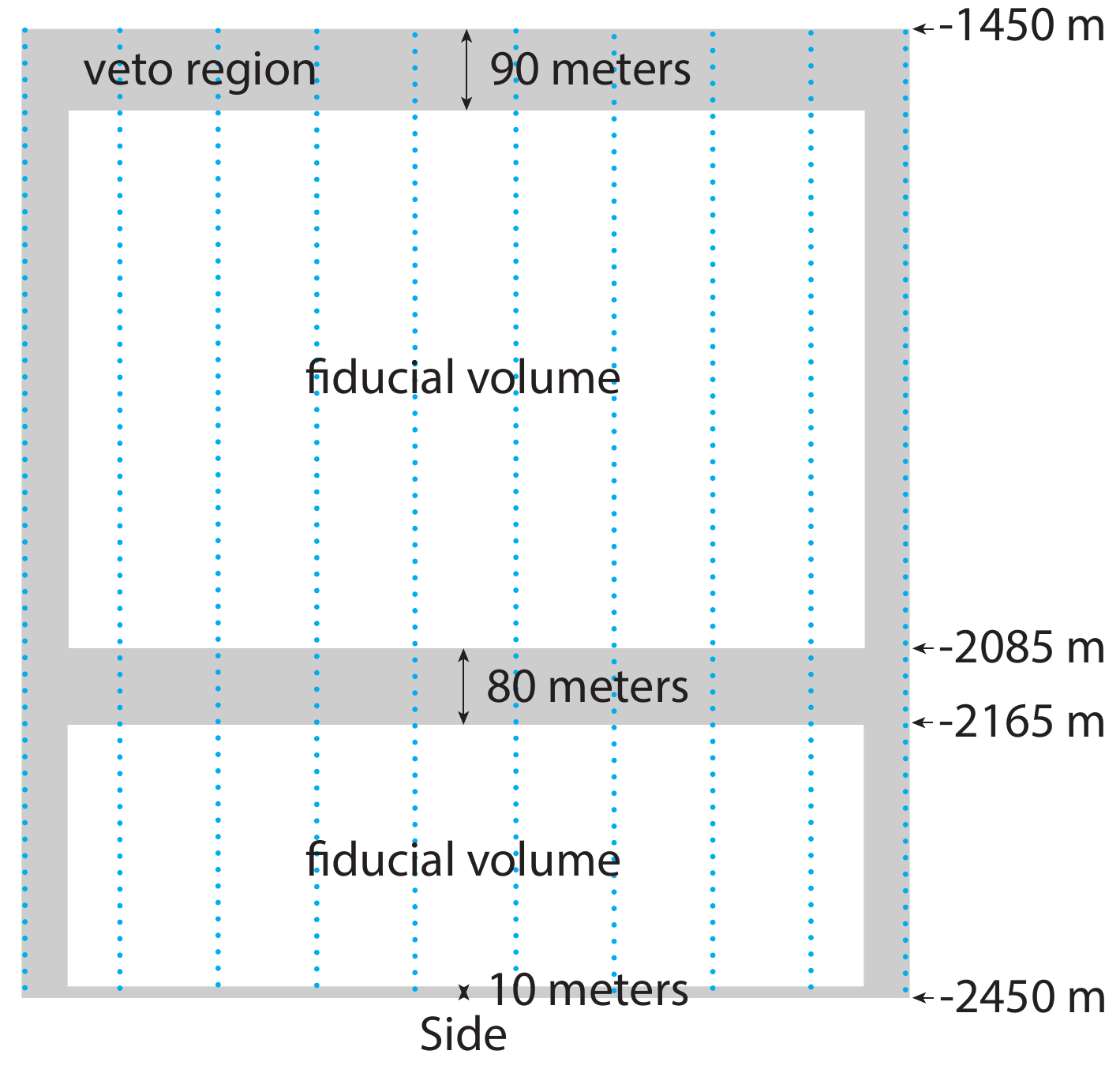}\,\,
\includegraphics[width=6 cm]{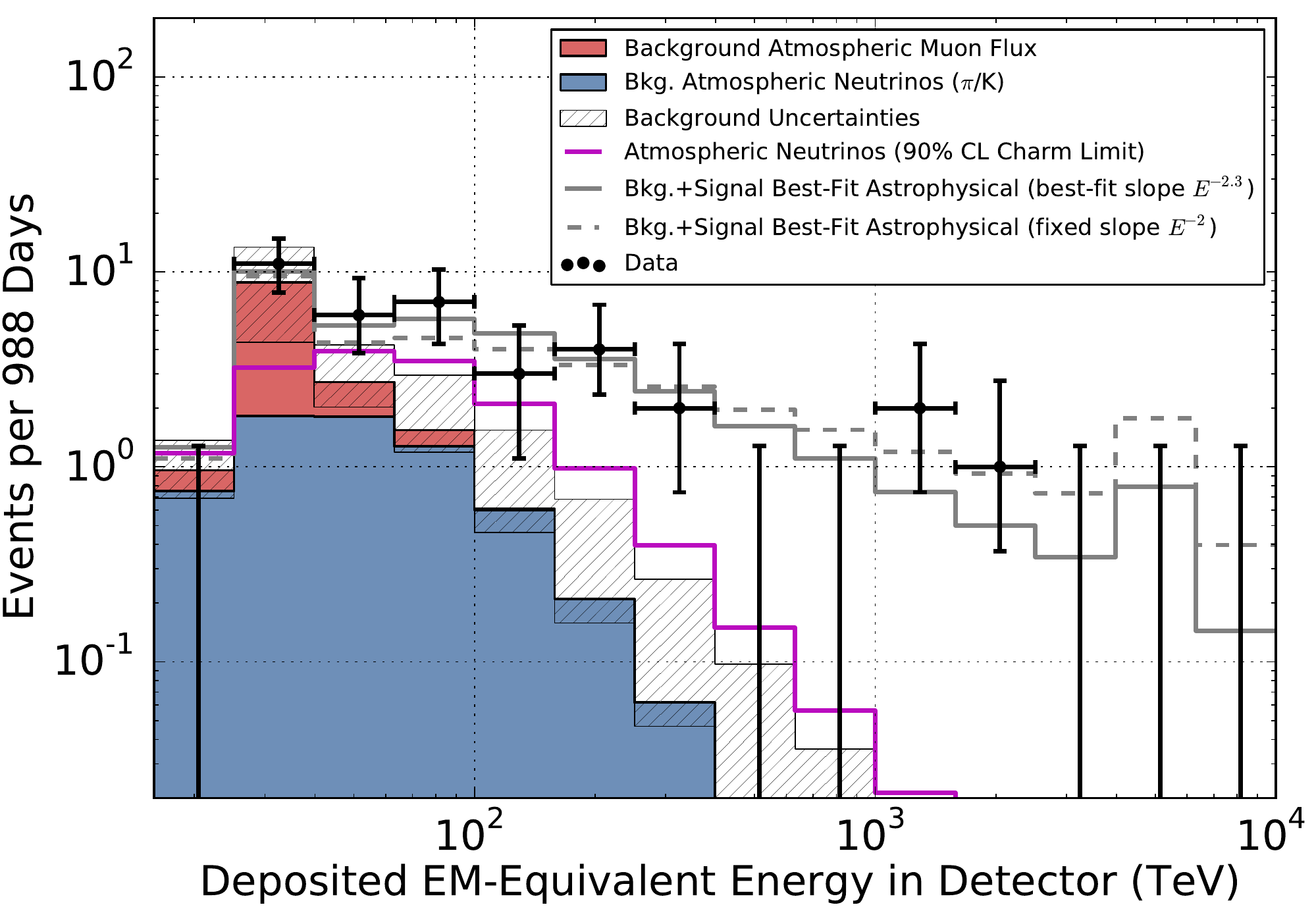}
  \caption{Left: Veto region for HESE analysis;  Right:
Energy distribution of events in the 3-year HESE analysis~\citep{2014PhRvL.113j1101A}.}
\label{fig8:veto}
\end{figure}

The veto region is shown in Fig.~\ref{fig8:veto} left along with the 
distribution in energy of events for data from three years.  The excess,
which has a significance of 5.7~$\sigma$~\citep{2014PhRvL.113j1101A}, is attributed
to high-energy neutrinos of astrophysical origin.
Assuming a differential spectral index of $-2$
for the astrophysical component and a flavor ratio at Earth of $1:1:1$, 
the astrophysical flux per flavor ($\nu+\bar{\nu}$) is 
\begin{equation}
E^2\phi(E) = 0.95\pm 0.3\times 10^{-8}\;{\rm GeV\, s}^{-1}{\rm sr}^{-1}{\rm cm}^{-2}.
\label{hard-spectrum}
\end{equation}
A fit to the astrophysical component without a prior constraint on its spectral
index allows spectral indexes from -2.0 to -2.3 depending on the background of
prompt neutrinos.  The best fit is at the lower boundary of the interval at
\begin{equation}
E^2\phi(E) = 1.5\times 10^{-8}\left(\frac{E}{100\,{\rm TeV}}\right)^{-0.3}\;{\rm GeV\, s}^{-1}{\rm sr}^{-1}{\rm cm}^{-2}.
\label{softer-spectrum}
\end{equation}
If the flavor ratio of anti-neutrinos at Earth 
is $(\bar{\nu}_e:\bar{\nu}_\mu:\bar{\nu}_\tau)=(1:1:1)$,
the harder spectrum (Eq.~\ref{hard-spectrum}) cannot continue unbroken above the
threshold of $6.3$~PeV for the Glashow process, $\bar{\nu}_e+e^-\rightarrow W^-$.
Three events with energies above 2 PeV would have been expected for
an unbroken $E^{-2}$ spectrum~\citep{2014PhRvL.113j1101A}.

A subsequent analysis extends the HESE selection strategy to lower energy by
defining a set of nested, increasingly smaller fiducial volumes~\citep{2015PhRvD..91b2001A}.
The analysis covers 641 days, overlapping the two-year HESE analysis~\citep{2013Sci...342E...1I}.
The lowest threshold corresponds to approximately 1 TeV of deposited energy in the detector.
The analysis finds 105 track events and 283 cascade events, including the two
1~PeV events seen in the two-year HESE analysis~\citep{2013Sci...342E...1I}.\footnote{It 
is important to keep in mind that the fixed threshold on deposited energy
biases against track events from charged current interactions of $\nu_\mu$ because
a significant fraction of the neutrino energy leaves the detector.  
Two recent analyses~\citep{2015PhRvL.114q1102A,2015arXiv150703991I} 
shows that the flavor ratios of the astrophysical neutrinos in IceCube are consistent
with the $1:1:1$ ratio expected at Earth after oscillations over astrophysical 
distances.} 
The analysis makes use of the different
characteristic energy and angular dependences of conventional atmospheric neutrinos,
prompt atmospheric neutrinos, muon background and astrophysical neutrinos to
fit the components separately.  
An 
upper limit of 1.5 times the prediction of \citet{2008PhRvD..78d3005E}
at 90\% confidence level is set on the charm contribution.  
The astrophysical component is fit with a relatively soft spectrum, 
\begin{equation}
E^2\phi(E) = 2.06_{-0.26}^{+0.35}\times 10^{-8}
\left(\frac{E}{100\,{\rm TeV}}\right)^{-0.46\pm0.12}\;{\rm GeV\, s}^{-1}{\rm sr}^{-1}{\rm cm}^{-2}.
\label{much-softer-spectrum}
\end{equation}
It is pointed out that such a soft spectrum of neutrinos, if produced by
proton-proton collisions in optically thin regions, would lead to 
$\pi^0\rightarrow \gamma \gamma$ production at a level greater than allowed by the
diffuse gamma-ray background observed by Fermi-LAT~\cite{2010PhRvL.104j1101A}. 
(See~\citet{2013PhRvD..88l1301M}).

  \begin{figure}[ht]
  \centering
\includegraphics[width=7 cm]{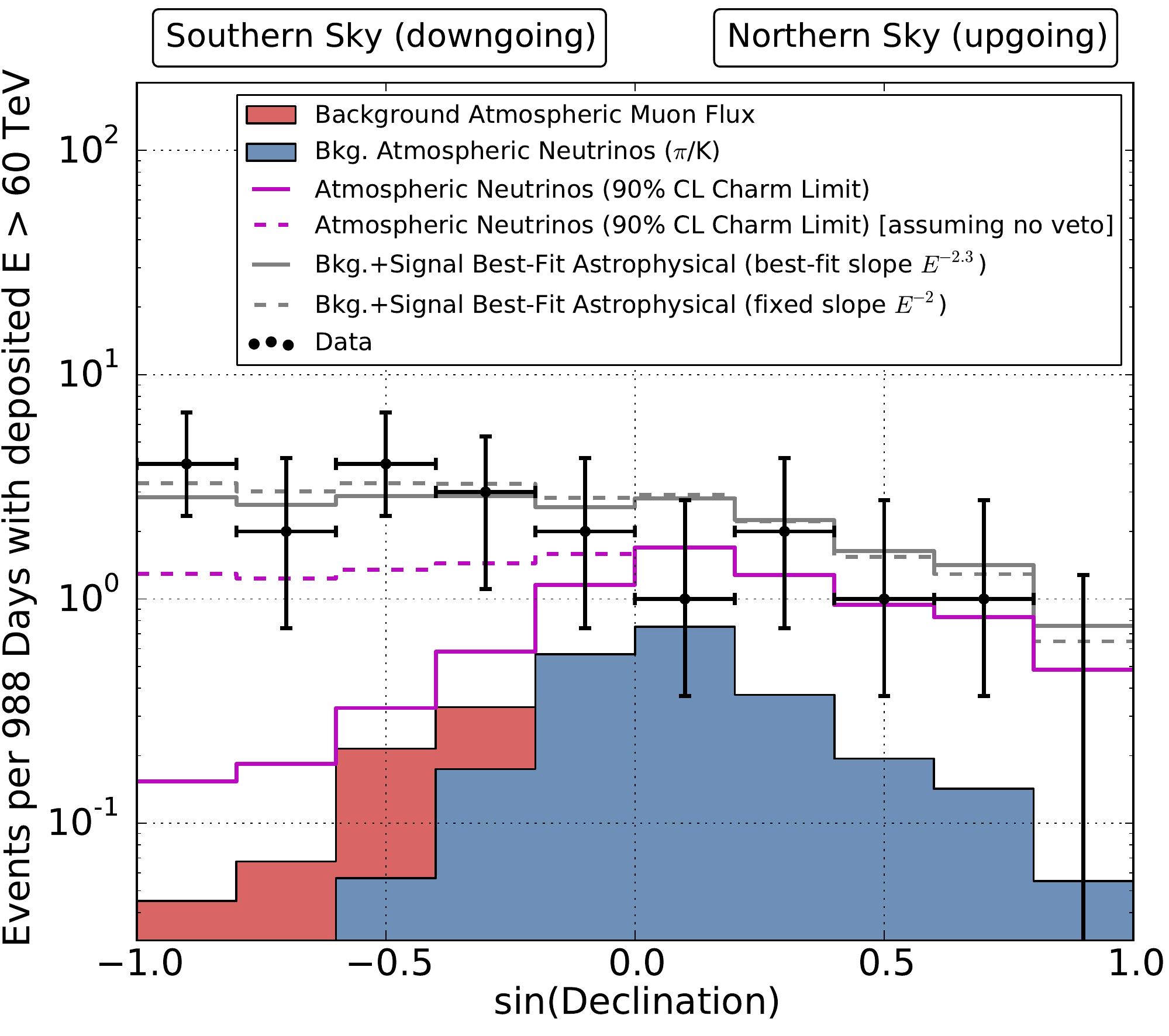}
  \caption{Angular distribution of events with deposited energy $>60$~TeV 
  in the HESE 3-year analysis~\citep{2014PhRvL.113j1101A}.  The solid pink histogram shows the 90\% c.l.
  upper limit from prompt neutrinos.  See text for discussion.}
  \label{fig9:angular}
\end{figure}

The prompt component of atmospheric neutrinos will have a spectral
index $\approx -2.7$, similar to that of the primary spectrum of nucleons
in the energy range of tens of TeV, approximately one power harder than
conventional atmospheric neutrinos from decay of kaons and pions.
It is therefore important to understand the low limits on the prompt
contribution to the background found in IceCube starting event analyses.
A feature of the atmospheric neutrino self-veto is relevant in this
context, as illustrated in Fig.~\ref{fig9:angular}.  
The solid pink histogram
shows the shape of angular distribution of prompt neutrinos from decay of charm
in the HESE analysis.  The broken pink histogram shows the prompt contribution
in the absence of the self-veto.   For neutrino energies below about 10 PeV,
neutrinos from decay of charmed hadrons are produced isotropically in the atmosphere.
In the absence of the neutrino self-veto, the expected flux would be nearly
isotropic apart from the suppression for nearly upward events due to
absorption in the Earth.  Instead, the neutrino self-veto suppresses
the downward contribution significantly.  The absence of a contribution
with this shape contributes to the low upper limit on charm from this~\citep{2015PhRvD..91b2001A} 
analysis.

  \begin{figure}[ht]
  \centering
\includegraphics[width=8 cm]{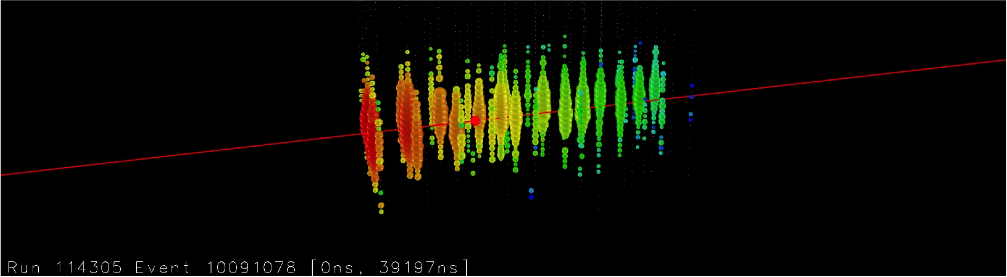}
  \caption{High-energy $\nu_\mu$-induced muon crossing IceCube from below the horizon~\citep{2013arXiv1311.7048T}.}
  \label{fig10:Schukraft}
\end{figure}
Figure~\ref{fig10:Schukraft} shows a neutrino-induced muon entering IceCube from
below the horizon in the analysis of data taken in 2009-2010~\citep{2013arXiv1311.7048T}.
With analysis of two more years of data (2010-2012) with the larger detector, the astrophysical signal is
becoming visible also in this channel at a level above 3$\sigma$~\cite{2015arXiv150704005I}. 
  The observed energy spectrum and fits of atmospheric and astrophysical
neutrino-induced muons are shown in Fig.~\ref{fig11:upnumu}.  The best fit parent astrophysical neutrino spectrum is
\begin{equation}
E^2\phi(E) = 1.7^{+0.6}_{-0.8}\times 10^{-8}\left(\frac{E}{100\,{\rm TeV}}\right)^{-0.2\pm 0.2}\;
{\rm GeV\,s}^{-1}{\rm sr}^{-1}{\rm cm}^{-2}.
\label{eq:astronumu}
\end{equation}

  \begin{figure}[t]
  \centering
\includegraphics[width=9 cm]{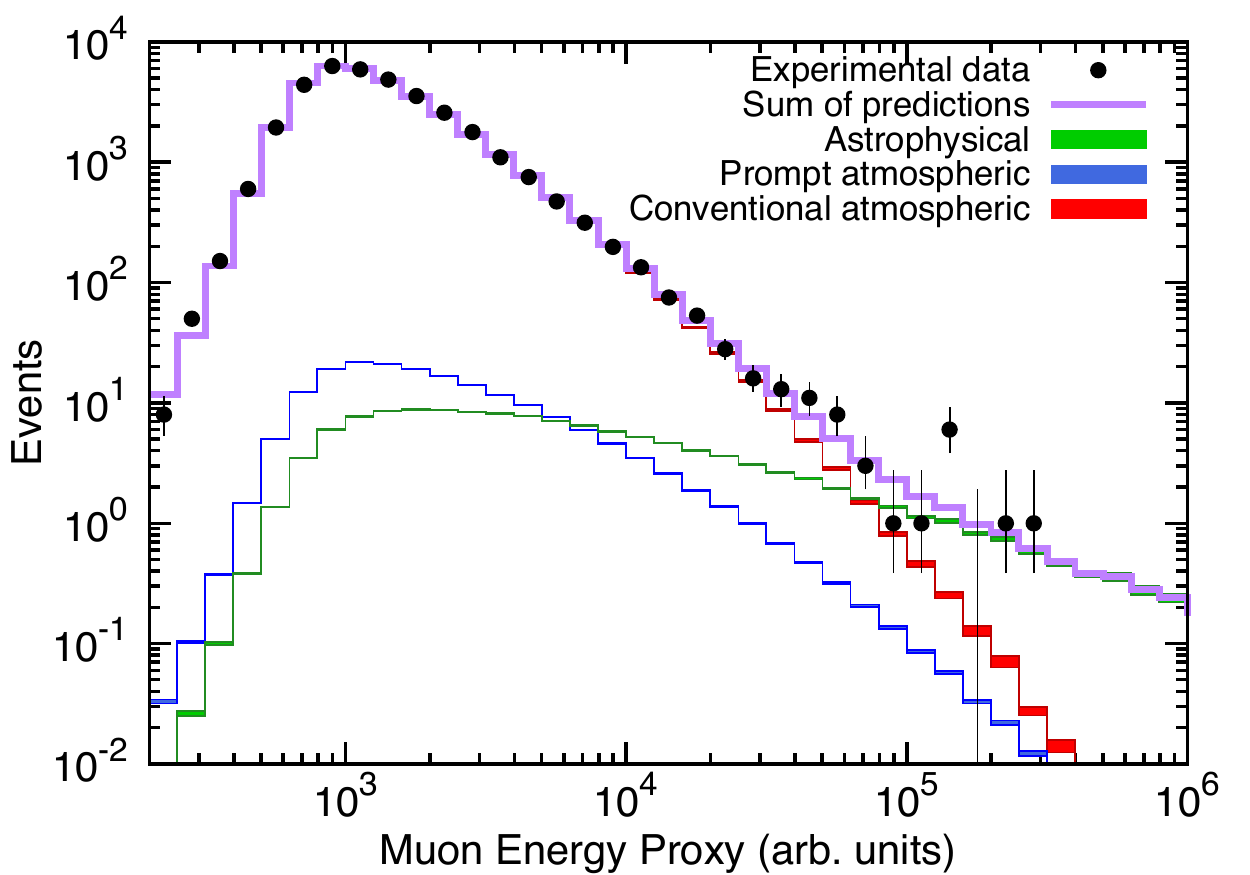}
  \caption{Energy spectrum of $\nu_\mu$-induced muons measured from 2010-2012~\cite{2015arXiv150704005I}.  }
  \label{fig11:upnumu}
\end{figure}

\section{Search for point sources}

Neutrino-induced muons provide the best sensitivity for point sources because
the muon tracks provide good angular resolution ($<1^\circ$) and the rates
are higher for a given flux because of the larger effective target volume
achieved by accepting muons that start outside the instrumented volume.
The most recently published search for steady sources~\citep{2014arXiv1406.6757I} 
covers four years, 2008-09 (IC40), 2009-10 (IC59), 2010-2011 ( IC79) and 
2011-12 (IC86).  No significant concentration of events is found.  

  \begin{figure}[ht]
  \centering
\includegraphics[width=8 cm]{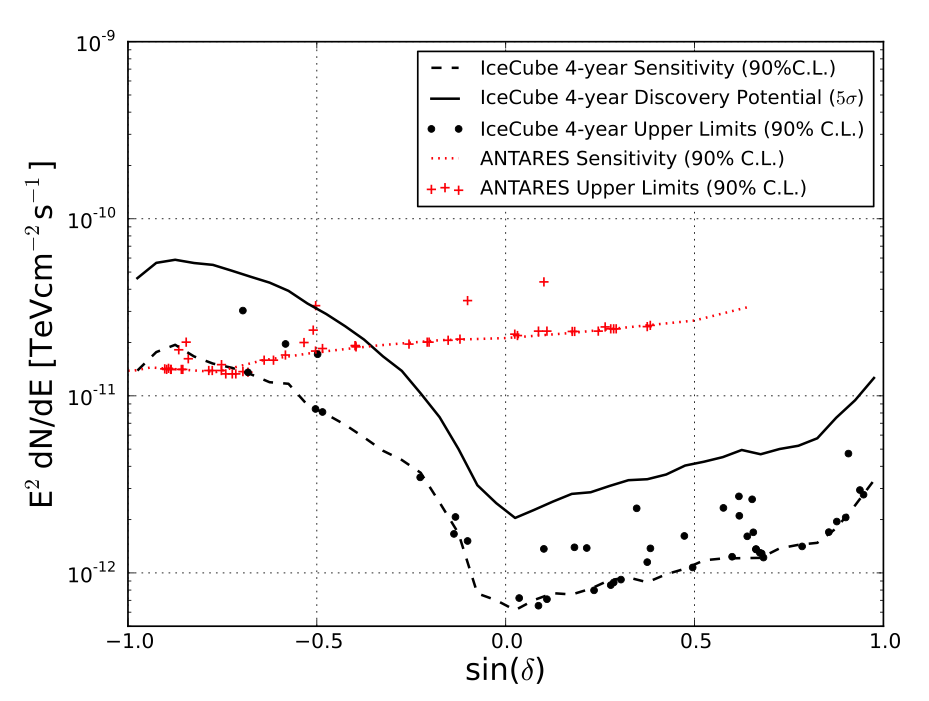}
  \caption{Upper limits on neutrinos from selected point sources from IceCube 
  (black) and ANTARES (red). (Figure from \citet{2014arXiv1406.6757I}.)}
  \label{fig12:ptsrc}
\end{figure}

In addition
to the all-sky scan, the analysis also looks at a selected list of 44 likely sources,
14 galactic and 30 extragalactic.  Upper limits for this search are shown
in Fig.~\ref{fig12:ptsrc} along with upper limits from 
ANTARES~\citep{2012ApJ...760...53A}.  The IceCube sensitivities and limits
are shown in this plot assuming $E^{-2}$ differential spectra.  In the
Northern sky, typical limits are at the level
\begin{equation}
E^2\frac{{\rm d}N}{{\rm d}E}\,\approx\,2\times 10^{-9}\,\,{\rm GeV\,cm}^{-2}{\rm s}^{-1}.
\label{eq:ptsrc}
\end{equation}
In order to
include the Southern sky, which is dominated by a high flux of atmospheric muons
at the South Pole, the energy threshold is set very high in IceCube to reduce the
background.\footnote{The small improvement in IceCube sensitivity for declinations
near $-90^\circ$ is achieved by using IceTop as a veto.}  ANTARES has good sensitivity
down to lower energy.  A joint analysis between ANTARES and IceCube is underway.
In addition, two more years of data (2012-14) with the full IceCube will soon be
added to the search for steady sources.

It is also of course possible to search for transient source of neutrinos, such as
AGN flares and gamma-ray bursts (GRBs).  Searches for events clustered in time
as well as events from particular sources known to flare have so far not
found any significant correlation~\citep{2015arXiv150300598A}.  
Particularly significant limits
come from the absence of neutrinos associated with GRBs~\citep{2012Natur.484..351A,2014arXiv1412.6510I}.  
The most recent IceCube search for neutrinos in association with GRB 
is inconsistent with standard models of optically thin gamma-ray bursts~\citep{2014arXiv1412.6510I}.
The analysis finds that no more than $\sim1$\% of the HESE flux 
consists of prompt emission from GRBs potentially
observable by exiting satellites.

  \begin{figure}[thb]
  \centering
\includegraphics[width=9 cm]{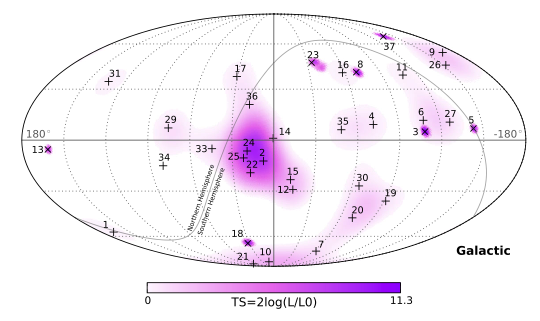}
  \caption{Map in galactic coordinates of the events from the three-year 
  HESE analysis~\citep{2014PhRvL.113j1101A}.}
  \label{fig13:HESEmap}
\end{figure}

The high-energy events from the HESE analysis are distributed over the sky, 
with some near the galactic plane (including a statistically insignificant cluster
near the galactic center), but with many far away from the plane.  
(See Fig.~\ref{fig13:HESEmap}.)  The angular
resolution for the cascade events is $\sim15^\circ$, so there are many potential
sources within the cone defined by each event.  
\citet{2014MNRAS.443..474P} show that it is nevertheless possible to
use the energetics of sources such as blazars and pulsar wind nebulae to select 
candidate sources within the error circles of many of the events with sufficient
power to be the corresponding sources.

Many, if not all, of the astrophysical events in the HESE sample 
are most likely of extragalactic origin.  
\citet{2014PhRvD..90d3005A} show that, in this case,
it is possible to use the observed luminosity density of the HESE flux (e.g. from
Eq.~\ref{hard-spectrum}) together with the upper limits from the point source search
(Eq.~\ref{eq:ptsrc}) to constrain the classes of sources responsible for the 
astrophysical neutrino flux in IceCube.  The argument is basically geometric, comparing
the integral over the neutrinos from all sources in the Universe 
with the flux from a typical nearby source~\citep{2008PhRvD..78h3011L}. 
Transient sources are constrained in
a similar way.  For steady sources the luminosity density is
\begin{equation}
L\rho = \frac{\rm energy}{{\rm source}\cdot{\rm time}}\times \frac{\rm sources}{\rm volume},
\label{eq:steady}
\end{equation}
while for transient sources (here $<100$~s) it is
\begin{equation}
L\rho = \frac{\rm energy}{\rm burst}\times \frac{\rm bursts}{{\rm volume}\cdot{\rm time}}.
\label{eq:transient}
\end{equation}

  \begin{figure}[thb]
  \centering
\includegraphics[width=9 cm]{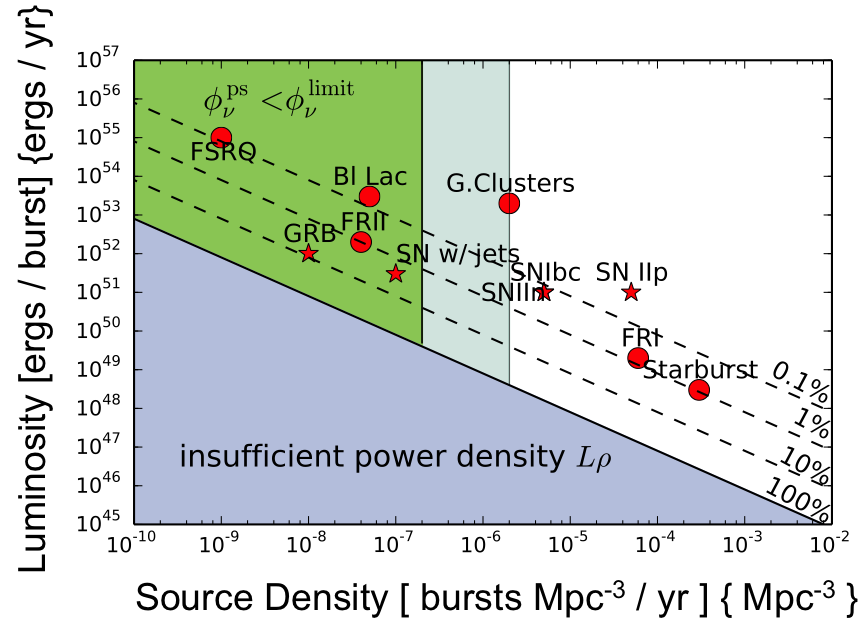}
  \caption{Diagram with limits on the contribution of various source
  classes to IceCube's astrophysical neutrino flux~\citep{2014arXiv1411.4385K}.
  Axis labels are in different brackets for [transient] and \{steady\} sources.}
  \label{fig14:Kowalski}
\end{figure}

Figure~\ref{fig14:Kowalski} by~\citet{2014arXiv1411.4385K} displays the present
IceCube constraints on various source classes on a single diagram.  The diagonal
line is obtained by equating the observed flux from all flavors 
($3\times$Eq.~\ref{hard-spectrum}) to
the integral of the neutrino flux from the Universe.  The result is
\begin{equation}
\xi\frac{L_\nu\rho R_H}{4\pi}= \frac{E_\nu^2{\rm d}N_\nu}{{\rm d}\Omega{\rm d}E_\nu}
\approx 2.8\times10^{-8}\,\frac{\rm GeV}{{\rm cm}^2{\rm sr\,s}},
\label{eq:Lrhointegral}
\end{equation}
where $R_H$ is the Hubble radius and $\xi$ is a cosmological factor of order $3$
which depends on the cosmological evolution of the particular class of sources.
This gives $L\rho\sim 10^{43}$~erg/Mpc$^3$/yr, which is the diagonal line
in the diagram.  A class of sources must have a value of luminosity density
greater than this to account for the observed flux.
The vertical lines represent the minimum density of sources (left line for steady sources)
or rate density (right line for transient sources) allowed by the non-observation
of the various source classes.  For example, equating the typical steady point source
limit from Eq.~\ref{eq:ptsrc} to the flux $L/(4\pi d_1^2)$ with $d_1=(4\pi\rho)^{-1/3}$
gives a minimum density of $\sim 10^{-7}$~Mpc$^{-3}$.

\vspace{-0.5cm}
\section{Plans for the future}

IceCube is currently in the position of having discovered high-energy astrophysical 
neutrinos without yet establishing what the sources are.  Upper limits are placing 
significant constraints on particular 
classes of potential sources.  For example, although blazars are 
attractive candidates~\citep{2014MNRAS.443..474P}, an analysis of
the Fermi-LAT catalog of blazars~\citep{2015arXiv150203104G} concludes that 
the sources in that catalog cannot
count for more that about 20\% of the observed astrophysical flux.  Starburst
galaxies~\citep{2006JCAP...05..003L} 
are an attractive potential class of sources, in part because of their
relatively low luminosity and relatively high density.  
On the other hand,
depending on how steeply the astrophysical spectrum extends 
to low energy~\citep{2015arXiv150104934S}, they may be in conflict with Fermi
observations of the diffuse gamma-ray background~\citep{2014arXiv1410.3696T}.

\vspace{-.5cm}
\subsection{Short term}

In this situation it is important to exploit multi-wavelength
and multi-messenger opportunities
as much as possible by making neutrino data available to other observers
as quickly as possible.  There are ongoing analyses looking for 
correlations of HESE events with existing catalogs
of bright transients, but the sensitivity of such an
analysis is determined by the threshold set in making the catalog.
Triggering followup observations with IceCube neutrinos instead
can significantly improve the chance of finding coincidences.
IceCube currently sends alerts under agreements
with various other observatories.  For example, the optical/X-ray follow-up (OFU/XFU)
program sends alerts to the Palomar Transient Facility and to SWIFT whenever two
neutrino-induced muons occur within a 100 seconds of each other a point
to the same direction within $3.5^\circ$.  The gamma follow-up (GFU) monitors 
selected sources and sends alerts to MAGIC and VERITAS.   

The OFU/XFU/GFU follow-up data stream
has recently been augmented to include single neutrinos of interest, which 
are being sent north at a rate of 5 mHz.  Events from the Southern hemisphere
are now included with criteria suitable for selecting likely
neutrino candidates, and H.E.S.S. is added to the list of receiving observatories.  
The single event
stream will be used to send events to the 
Astrophysical Multimessenger Observatory Network (AMON)~\citep{2013APh....45...56S}
for sharing of sub-threshold data among multi-messenger observatories.

For HESE events, a starting event filter has been implemented that will
provide alerts in real time.  A short message with information from
the online filter, including a probability of track vs. cascade,
will be sent to AMON.  
The treatment of the starting events will depend on their brightness.
GCN alerts will be generated for events with more than 6000 p.e. ($\sim 30$~TeV and 10-15/year).  
Starting events with more than 1500 but less than 6000 p.e. ($\approx 4/$day) will be
included in the AMON sub-threshold data sharing scheme.
Upgrades to the satellite connection from the South Pole will allow
full event information to be sent with delays of order one minute.
This will provide the basis for full reconstruction of events.  Alerts
can be revised accordingly for the brightest events.

\subsection{Long term}
\vspace{-.5cm}
\noindent
For the longer term future, plans for expanding IceCube to collect more
high energy neutrinos are underway~\citep{2014arXiv1412.5106I}.  
Figure~\ref{fig15:Gen2} is a picture of
what such an expanded array might look like.  IceCube Gen2 includes the
Precision IceCube Next Generation Upgrade
(PINGU)~\citep{2014arXiv1401.2046T} for precision studies of neutrino
physics, including mass hierarchy.

  \begin{figure}[th]
  \centering
\includegraphics[width=8 cm]{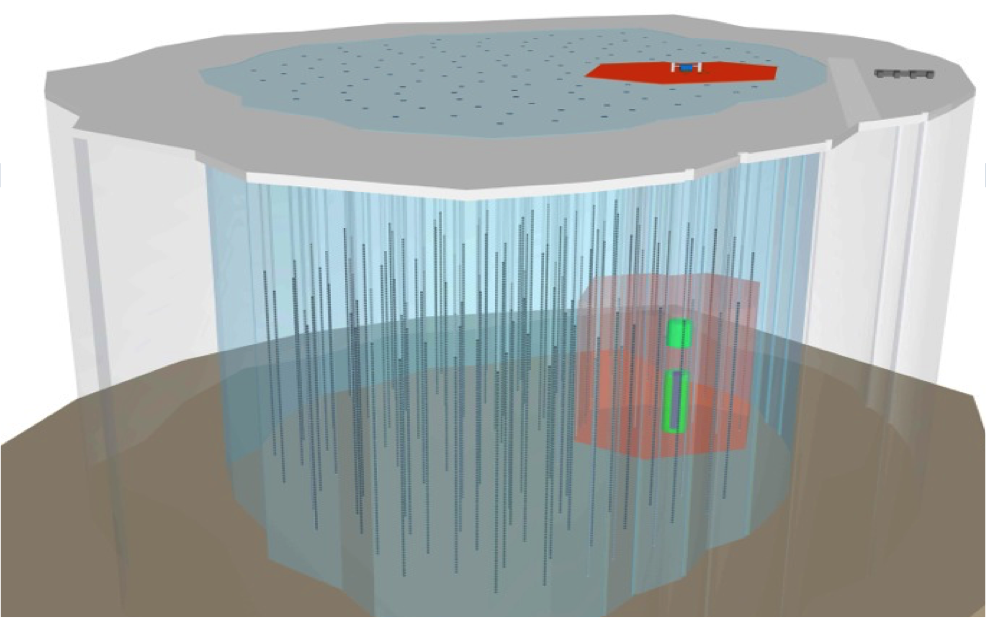}
  \caption{Concept for an expanded version of IceCube~\citep{2014arXiv1412.5106I}.
  PINGU is indicated by the dense shading inside DeepCore.}
  \label{fig15:Gen2}
\end{figure}

\vspace{.5cm}
\begin{acknowledgement}

I thank the HAWC Collaboration for the opportunity to bring congratulations
from IceCube on the occasion of the inauguration of HAWC.
I am grateful to Erik Blaufuss and Marek Kowalski for help preparing this paper,
and I thank my colleagues in IceCube for the science summarized in this paper.  It is a
pleasure to acknowledge the National Science Foundation for their support
of IceCube and of my own research.  A full list of agencies supporting 
IceCube is posted at http://icecube.wisc.edu/Collaboration/funding.
The list of IceCube institutions is available at http://icecube.wisc.edu/collaboration.

\end{acknowledgement}

\end{document}